\documentclass[sigconf]{acmart}
\usepackage[utf8]{inputenc}
\usepackage{adjustbox}
\usepackage[ruled,vlined]{algorithm2e}
\usepackage{algorithmic}
\usepackage{graphicx}
\usepackage{textcomp}
\usepackage{xcolor}
\usepackage{balance}
\usepackage{enumitem}
\usepackage{setspace}
\usepackage{wrapfig}
\usepackage{listings}
\usepackage{color}
\usepackage{caption}
\usepackage{subcaption}
\usepackage{balance}

\newcommand{\etal}{\emph{et al. }}
\newcommand{\ie}{\emph{i.e.}, }

\newcommand{\HAS}{\emph{HTTP Adaptive Streaming }}

\newcommand{\DCT}{\emph{Discrete Cosine Transform }}

\renewcommand{\footnotesize}{\fontsize{7pt}{7pt}\selectfont}

\copyrightyear{2023}
\acmYear{2023}
\setcopyright{rightsretained}
\acmConference[GMSys '23]{First International ACM Green Multimedia Systems Workshop}{June 7--10, 2023}{Vancouver, BC, Canada}
\acmBooktitle{First International ACM Green Multimedia Systems Workshop (GMSys '23), June 7--10, 2023, Vancouver, BC, Canada}
\acmDOI{10.1145/3593908.3593942}
\acmISBN{979-8-4007-0196-2/23/06}

\begin{document}
\title{Green Video Complexity Analysis for Efficient Encoding in Adaptive Video Streaming}
\author{Vignesh V Menon$^{1}$, Christian Feldmann$^{2}$, Klaus Schoeffmann$^{1}$, Mohammad Ghanbari$^{3}$, and Christian Timmerer$^1$}
\affiliation{
  \institution{\small$^1$Christian Doppler Laboratory ATHENA, Institute of Information Technology (ITEC), Alpen-Adria-Universität Klagenfurt, Austria}
  \institution{\small$^2$Bitmovin, Austria}
  \institution{\small$^3$School of Computer Science and Electronic Engineering, University of Essex, Colchester, UK}
}
\renewcommand{\shortauthors}{Vignesh V Menon~\etal}
\begin{abstract}
For adaptive streaming applications, low-complexity and accurate video complexity features are necessary to analyze the video content in real time, which ensures fast and compression-efficient video streaming without disruptions. State-of-the-art video complexity features are Spatial Information (SI) and Temporal Information (TI) features which do not correlate well with the encoding parameters in adaptive streaming applications. To this light, Video Complexity Analyzer (VCA) was introduced, determining the features based on \DCT~(DCT)-energy. This paper presents optimizations on VCA for faster and energy-efficient video complexity analysis. Experimental results show that VCA v2.0, using eight CPU threads, Single Instruction Multiple Data (SIMD), and low-pass DCT optimization, determines seven complexity features of Ultra High Definition 8-bit videos with better accuracy at a speed of up to 292.68 fps and an energy consumption of 97.06\% lower than the reference SITI implementation.
\end{abstract}



\maketitle

\section{Introduction}
Video complexity analysis is a critical step for numerous video streaming applications. Haseeb~\etal~\cite{rd_siti_ref} used spatial and temporal complexity information in rate-distortion modeling.
The video complexity evaluation is also essential in QoE evaluation metrics~\cite{vqa_siti_ref, qoe_siti_ref2, qoe_siti_ref3, qoe_siti_ref1, vqa_mhv_ref, vqa_icip_ref}. Pinson~\etal~\cite{vqa_siti_ref} measured the video quality objectively by utilizing the spatial content of the sequences. Barman~\etal~\cite{qoe_siti_ref2}, Goring~\etal~\cite{qoe_siti_ref3}, and Zadtootaghaj~\etal~\cite{qoe_siti_ref1} proposed machine learning-based QoE models, where spatial and temporal complexity values are used along with other influential factors for quality estimation of gaming videos. 
Fast video complexity analysis is used in online per-title encoding schemes~\cite{de_cock_complexity-based_2016}, which determine optimized resolution~\cite{opte_ref}, bitrate-ladder~\cite{ppte_ref}, framerate~\cite{coda_ref}, compressibility~\cite{compressibility_ref} and other relevant encoding parameters for live \HAS~(HAS) applications~\cite{jtps_ref}.

\begin{figure}[t]
    \centering
    \includegraphics[trim={0.0cm 11.5cm 29.50cm 0cm}, clip,width=0.80\linewidth]{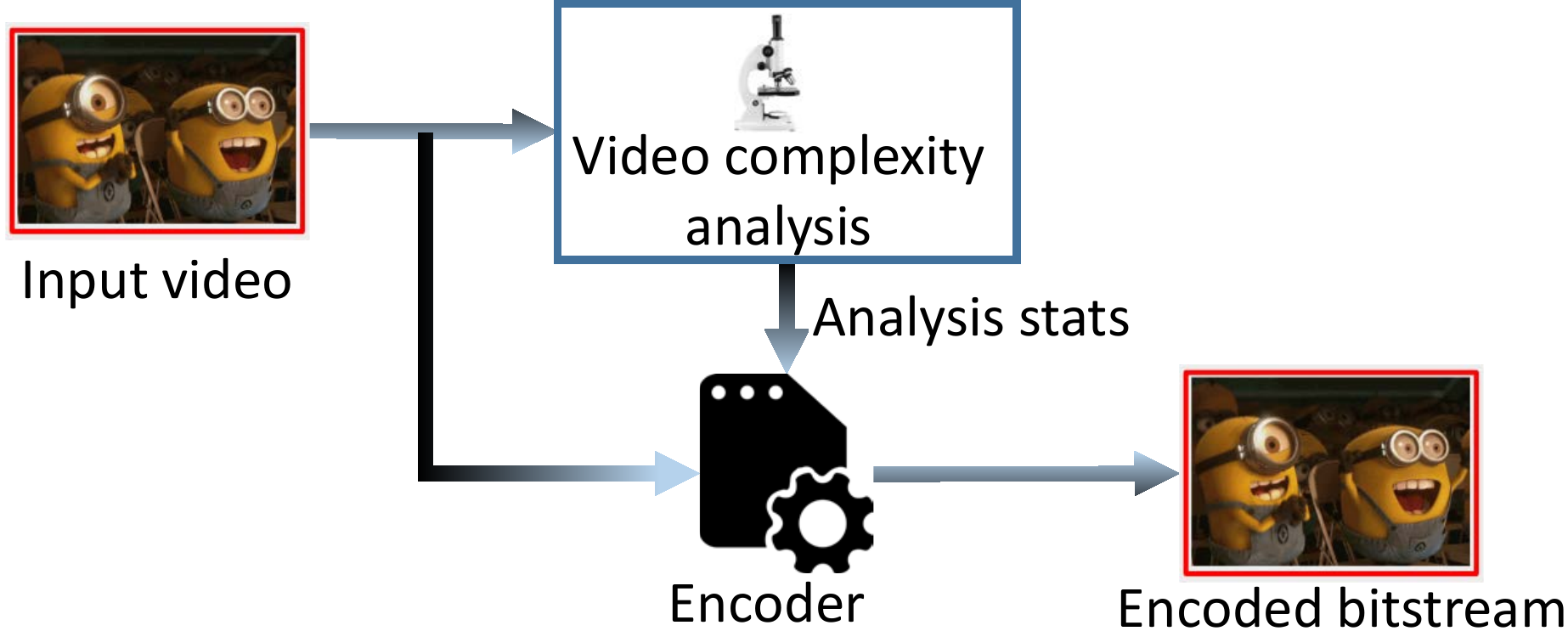}
\vspace{-0.85em}    
\caption{Content-adaptive encoding framework using video content complexity analysis.}
\vspace{-1.2em}
\label{fig:cae_vca}
\end{figure}

SITI\footnote{\label{siti_ref}\href{https://github.com/Telecommunication-Telemedia-Assessment/SITI}{https://github.com/Telecommunication-Telemedia-Assessment/SITI}, last access: Apr 20, 2023.} is the state-of-the-art open-source software to evaluate spatial and temporal complexity. Spatial Information (SI) indicates the maximum spatial detail in a video. Each video frame $F(x,y,t)$ is filtered by the Sobel filter $[Sobel(F(x,y,t))]$ and the standard deviation for each Sobel-filtered frame $std[Sobel(F(x,y,t))]$ is then calculated. The maximum std value across the video sequence is chosen to be the SI which is represented as $SI = max\{std[Sobel(F(x,y,t))]\}$. Temporal Information (TI) is the maximum temporal variation between two successive frames. The difference between successive frames is represented as $D(x,y,t) = F(x,y,t)) - F(x,y,t-1))$. The standard deviation $std[D(x,y,t)]$ of each difference $D(x,y,t)$ is calculated. TI is calculated as the maximum value of $std[D(x,y,t)]$ across the video sequence, represented as $TI = max\{std[D(x,y,t)] \}$.

\section{Video complexity analyzer}
Video complexity analyzer\footnote{\label{ref_vca}\href{https://vca.itec.aau.at}{https://vca.itec.aau.at}, last access: Apr 20, 2023.} (VCA) determines seven blockwise DCT-energy-based features, the luma texture energy $E_{Y}$, the gradient of the luma texture energy $h$, the luminescence $L_{Y}$, the chroma texture energy $E_{U}$ and $E_{V}$ (for U and V planes), and the chrominance $L_{U}$ and $L_{V}$ (for U and V planes)~\cite{vca_ref}. For a green and fast video complexity analysis, the following optimizations are introduced: 

\textit{(1) Multi-threading optimization:}
To optimize the performance in multi-core CPUs, VCA leverages the multi-threading mechanism. Multi-threading optimization creates multiple threads within a VCA execution instance,  which executes independently but concurrently, sharing process resources. Independent threads carry out DCT-energy computation per block.

\textit{(2) x86 SIMD optimization:} VCA leverages the SIMD optimization~\cite{x86_simd_ref} of DCT functions implemented as intrinsic and assembly codes for x86 architecture. The intrinsic code of DCT is executed for Intel SSSE3 architecture. Handwritten Intel SSSE3 and AVX2 instructions accelerate the DCT calculation per block.

\begin{figure}
\centering
    \subfloat[]{\includegraphics[width=0.43\linewidth]{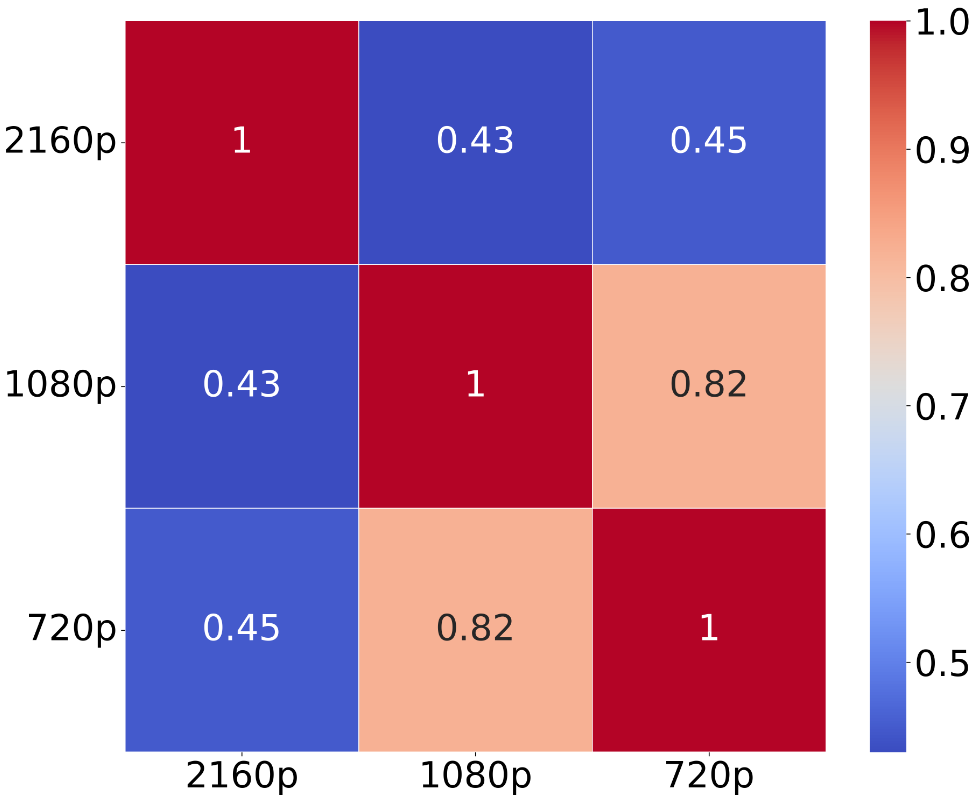}}
    \hfill
    \subfloat[]{\includegraphics[width=0.43\linewidth]{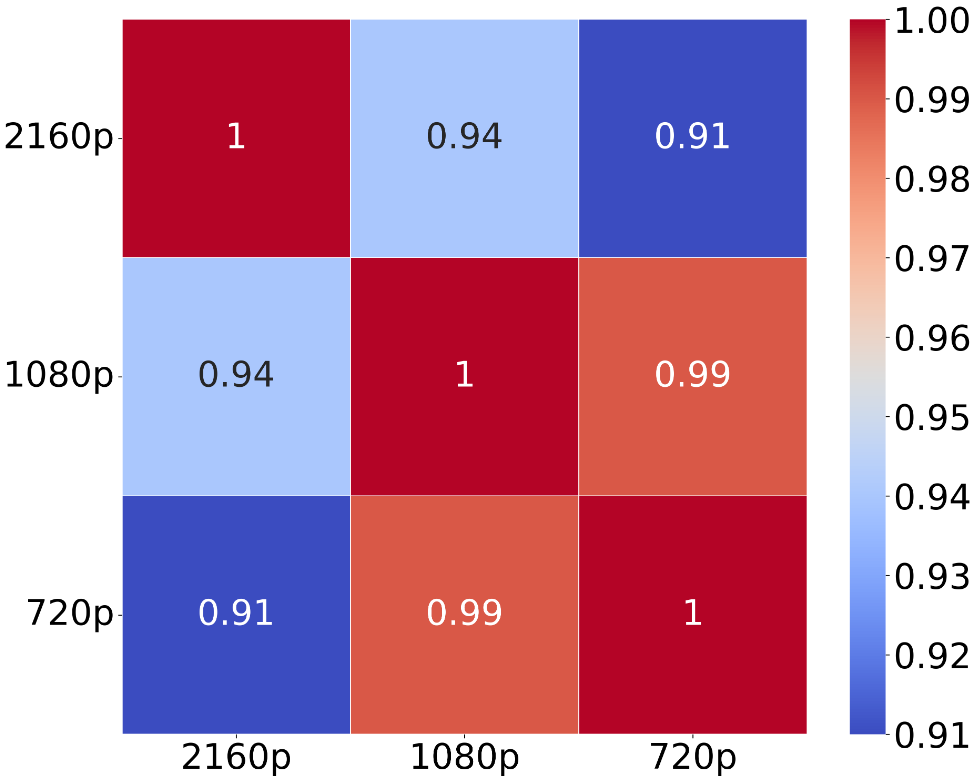}}
\vspace{-0.75em}
\caption{PCC between the spatial complexity features (a) SI and (b) $E_Y$ across multiple resolutions for the VCD dataset~\cite{VCD_ref}.}
\vspace{-0.75em}
\label{fig:si_e_res_corr}
\end{figure}

\textit{(3) Low-pass DCT optimization:} Unlike SI, the $E_Y$ feature exhibits better correlation across resolutions as shown in Figure~\ref{fig:si_e_res_corr}. Thus, VCA v2.0 implements a low-pass DCT optimization, where the complexity features are evaluated on the video spatially downsampled by a factor of two. This optimization is expected to increase the processing speed and lower the energy consumption.

\begin{figure}[t]
    \centering
    \begin{subfigure}{0.2\textwidth}
    \centering    
    \includegraphics[width=0.93\textwidth]{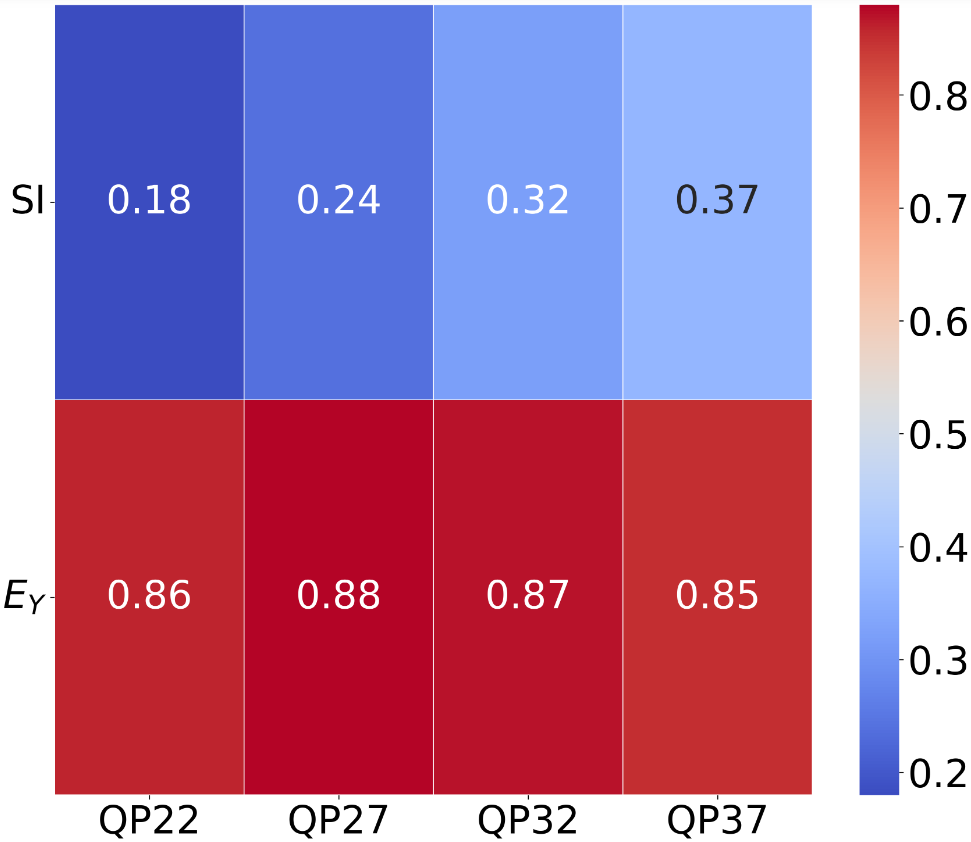}
\caption{}
\label{fig:all_intra_corr}
\end{subfigure}
\hfill
\begin{subfigure}{0.2\textwidth}
    \centering
    \includegraphics[width=0.9\textwidth]{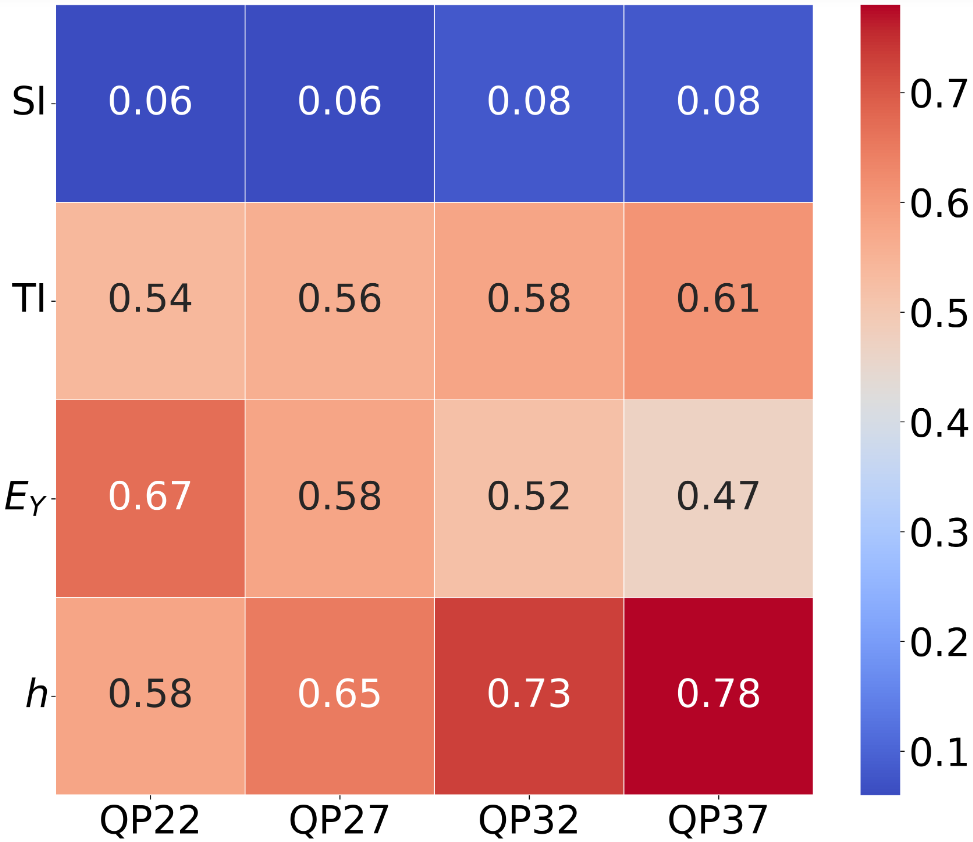}
\caption{}
\label{fig:bitrate_inter_corr_x265}
\end{subfigure}
\vspace{-0.85em}
\caption{PCC between (a) SI and $E_{Y}$, respectively, with bitrate in All Intra configuration with \textit{medium} preset of x265 encoder and (b) SI, $E_Y$, TI and $h$ with \textit{bitrate} in the Low Delay P picture configuration with \textit{ultrafast} preset of x265 encoder for the VCD dataset~\cite{VCD_ref}.}
\vspace{-1.3em}
\end{figure}

\section{Performance Evaluation}
First, the accuracy of $E_Y$ is compared to the state-of-the-art SI feature. In this light, the correlation of SI and $E_Y$ features with the bitrate in All Intra configuration~\cite{boyce_jvet-j1010_2018} is evaluated since the bitrate in that configuration is considered as the ground truth of the spatial complexity. As shown in Figure~\ref{fig:all_intra_corr}, the average Pearson Correlation Coefficient (PCC) of SI with bitrate is 0.28, while the average PCC of $E_Y$ with bitrate is 0.86, respectively. Thus, $E_Y$ represents spatial complexity better than SI.

\begin{figure}[t]
    \centering
    \begin{subfigure}{0.47\textwidth}
    \centering    
    \includegraphics[width=0.95\textwidth]{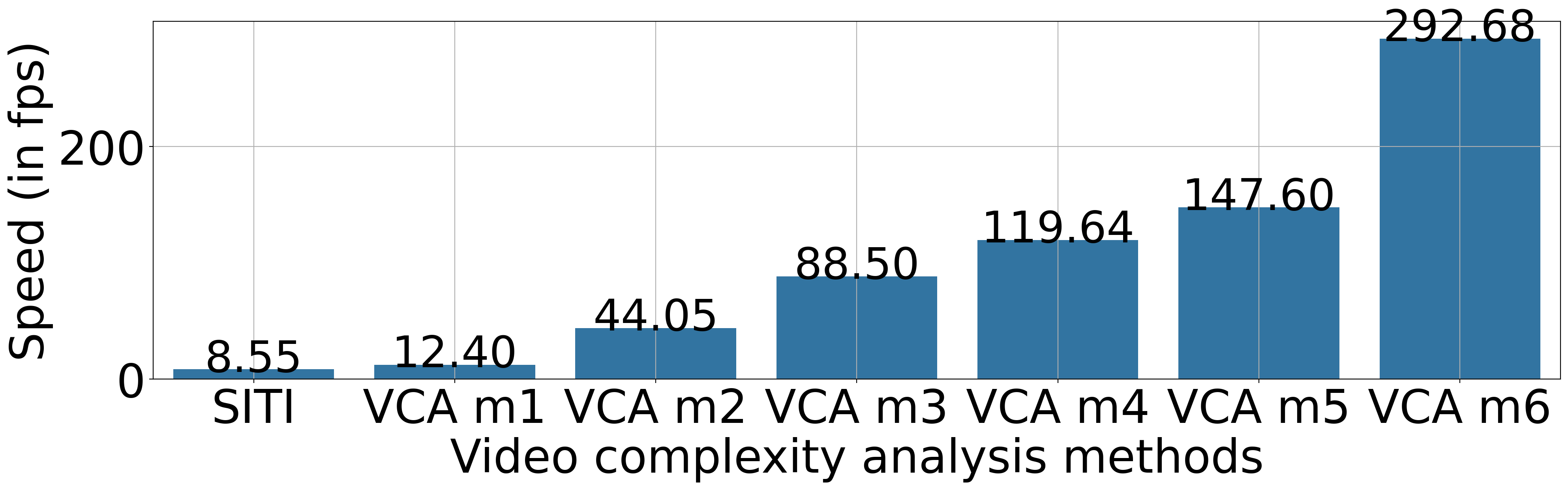}
\caption{}
\label{fig:time_vca}
\end{subfigure}
\begin{subfigure}{0.47\textwidth}
    \centering
    \includegraphics[width=0.95\textwidth]{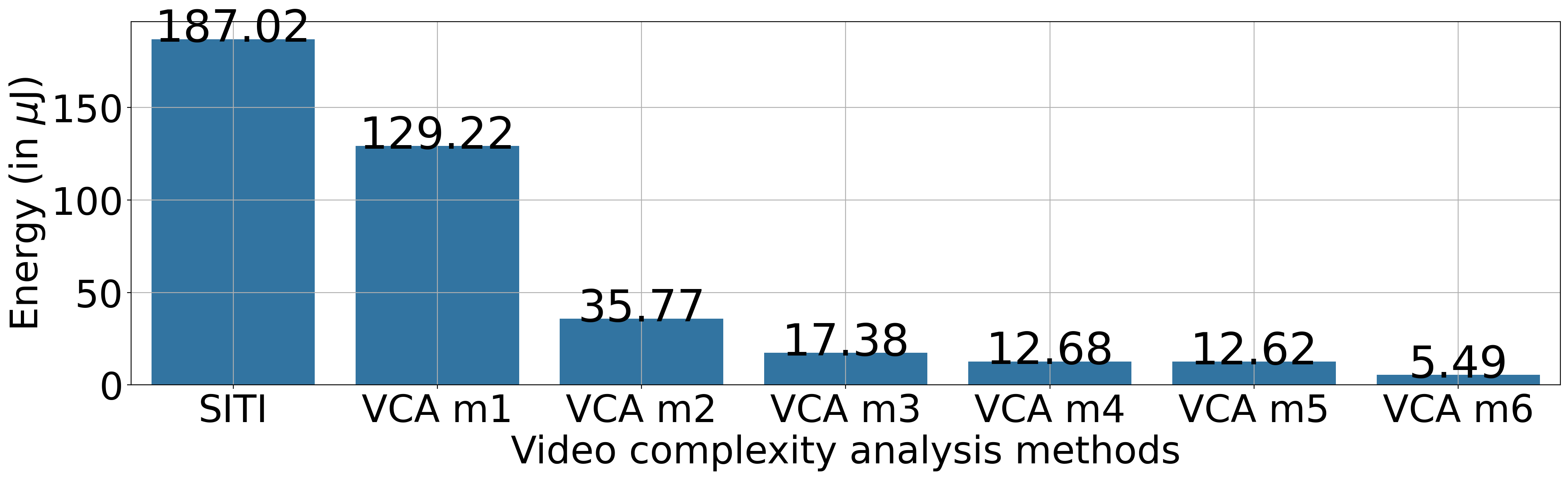}
\caption{}
\label{fig:energy_vca}
\end{subfigure}
\vspace{-0.85em}
\caption{(a) Processing time and (b) energy consumption of video content complexity analysis methods.}
\vspace{-1.0em}
\label{fig:vca_comp_plots}
\end{figure}

Second, the accuracy of $E_Y$ and $h$ are evaluated in terms of their correlation with the encoding bitrate in the Low Delay P picture configuration~\cite{boyce_jvet-j1010_2018}. As shown in Figure~\ref{fig:bitrate_inter_corr_x265}, $E_Y$ and $h$ strongly correlate with the encoding bitrate.

Third, the analysis speed and energy consumption of VCA are compared to SITI for Ultra High Definition (UHD) 8-bit videos on a computer with an Intel i7-11370H processor and 16GB RAM. The energy consumption is measured using \textit{codecarbon}\footnote{\href{https://codecarbon.io/}{https://codecarbon.io/}, last access: Apr 20, 2023}.
As shown in Figure~\ref{fig:vca_comp_plots}, the reference SITI implementation analyzes content at a rate of 8.55 fps with an energy consumption of 187.02 $\mu$J. The reference VCA implementation (represented as VCA m1), \ie without multi-threading, SIMD, and low-pass optimizations, yields 12.40 fps with energy consumption of 30.90\% lower than the SITI implementation. Using only low-pass DCT optimization (VCA m2), the energy consumption decreases by 72.01\% compared to VCA m1. Adding SIMD optimization to VCA m2 (represented as VCA m3), energy consumption drops by 51.41\%. VCA m4, m5, and m6 modes represent using two, four, and eight CPU threads, respectively. As the number of threads used increases, the analysis speed increases while the energy consumed decreases. With all optimizations enabled and using eight CPU threads, VCA is 33 times faster than SITI, with 97.06\% lower energy consumption.       
\section{Conclusions}
This paper presents an energy-efficient video content complexity analysis for efficient encoding in adaptive video streaming applications. Low-complexity DCT-based energy features are extracted using VCA v2.0, which encoders use to derive decisions like bitrate-ladder, frame-type, block-partitioning, and much more. Multi-threading, x86 SIMD, and low-pass DCT optimizations improve the energy efficiency of the VCA implementation. Compared to the state-of-the-art SITI implementation of video complexity analysis, VCA v2.0 yields a better estimation of video complexity, with an energy consumption reduction of 97.06\%. VCA is published under the GNU GPLv3 license at \href{https://github.com/cd-athena/VCA}{https://github.com/cd-athena/VCA}.    

\section{Acknowledgment}
The financial support of the Austrian Federal Ministry for Digital and Economic Affairs, the National Foundation for Research, Technology and Development, and the Christian Doppler Research Association is gratefully acknowledged. Christian Doppler Laboratory ATHENA: \url{https://athena.itec.aau.at/}.

\balance
\bibliographystyle{ACM-Reference-Format}
\bibliography{acmart.bib}
\end{document}